# A machine-learning based closed orbit feedback for the SSRF storage ring

Ruichun Li[1,2,3], Qinglei Zhang[1,4,*], Bocheng Jiang[5,†], Zhentang Zhao[1,2,3,4], Changliang Li[1,4], Kun Wang[1,4], and Dazhang Huang[1,4]

[1] *Shanghai Synchrotron Radiation Facility, Shanghai Advanced Research Institute, Chinese Academy of Sciences, Shanghai 201210, China*
[2] *Shanghai Institute of Applied Physics, Chinese Academy of Sciences, Shanghai 201800, China*
[3] *School of Physical Science and Technology, ShanghaiTech University, Shanghai 201210, China*
[4] *University of Chinese Academy of Sciences, Beijing 100049, China*
[5] *Chongqing University, Chongqing 401331, China*

In order to improve the stability of synchrotron radiation, we developed a new method of machine learning-based closed orbit feedback, and piloted it in the storage ring of the Shanghai Synchrotron Radiation Facility (SSRF). In our experiments, not only can the machine learning-based closed orbit feedback carry out horizontal, vertical and RF frequency feedback simultaneously, but it also has better convergence and convergence speed than the traditional Slow Orbit Feed Back (SOFB) system. What's more, the residual values of the correctors' currents variations after correction can be almost ignored. This machine learning-based new method is expected to establish a new closed orbit feedback system and improve the orbit stability of the storage ring in daily operation.

[*] Corresponding author at: Shanghai Synchrotron Radiation Facility, Shanghai Advanced Research Institute, Chinese Academy of Sciences, Shanghai 201210, China
E-mail addresses: zhangqinglei@zjlab.org.cn (Qinglei Zhang)
[†] Corresponding author at: Chongqing University, Chongqing 401331, China
E-mail addresses: jiangbocheng@cqu.edu.cn (Bocheng Jiang)

# I. INTRODUCTION

Synchrotron light source is one of the most powerful tools in modern science and technology. It was already developed from 1st-generation to 3rd-generation to provide highly brilliant and highly stable synchrotron radiation to fulfill the advanced experimental conditions in frontier research[1]. For 3rd generation synchrotron light sources, it is important to have a highly stable electron beam in order to achieve highly brilliant and highly stable synchrotron radiation. This includes orbit stability, current stability, control of the collective effects and energy stabilization[2-5]. Among them, the orbit stability is essential and can be further improved with a machine learning-based closed orbit feedback, which is under exploring in recent years.

As an advanced 3rd-generation light source, Shanghai Synchrotron Radiation Facility (SSRF) is based on a 3.5 GeV storage ring with an average beam current of more than 200 mA[6]. The SSRF initial storage ring consists of 20 standard double bend achromatic (DBA) cells with four super-periods, but its lattice was upgraded with two DBA cells with high magnetic field super-bends. In addition, the upgrades also include constructing two double-mini-$βy$ optics (DMB), adding a third harmonic superconducting radio frequency (SRF) cavity module, installing a superconducting wiggler (SCW) and 13 new undulators[7]. The number of beamlines and experimental stations has increased to 32 and 53 respectively after the upgrade. Therefore, higher and critical requirements are put forward for the stability issues.

During the design stage of a storage ring, considerable efforts must be made to optimize its performance parameters, such as tune, chromaticity, natural beam emittance and dynamic aperture. In an ideal state, the storage ring parameters are operated in the design values and the beam moves around a closed orbit in the storage ring. However, the actual situation is often different from the theoretical model. Not only will magnetic field errors and magnet misalignments cause closed orbit distortion(COD), but also thermal expansion and contraction, settlement deformation of the ground and other factors can cause various displacements and mechanical vibrations of the magnets in the storage ring, leading to the COD[8].

We can classify these changes into slow and fast changes according to their variation time scales that cause COD. Among them, settlement of the ground is in a slow change and is typically measured in years, which can be overcome by periodical re-alignment. Temperature change is also classified as a slow change, measured in hours. Temperature change generally refers to the change of ambient temperature, which causes the phenomenon of thermal expansion and contraction, resulting in the drift of the magnets and beam position monitors (BPM), and finally causing the COD, which is a major challenge for the orbit stability. Different amplitudes of vibration can be generated by mechanical components of the storage ring due to facility utilities and human activities. Located in the urban area, SSRF suffers from significant ground vibration from the surrounding traffic environment. The magnet power supply ripples will lead to the beam orbit motion, which are also classified as fast changes[9].

Furthermore, the orbital changes near photon source points will directly cause the change of the position and angle of the synchrotron radiation, which affects the results and quality of user experiments. Therefore, it is essential to adopt an effective method to reduce and control the COD, improve the beam stability and ultimately ensure a stable operation of synchrotron light source.

Studies on possible schemes for stabilizing the beam in a storage were carried out as early as 1970s and 1980s[10,11]. In a storage ring, beam position stabilization systems can be classified into global feedback and local feedback systems. The global feedback systems attempt to stabilize the beam orbit at all the concerned position in a storage ring, while the local feedback systems only affect one or a few adjacent beamlines. The common local feedback systems use the local bump method and this method only changes the orbit in an adjacent area, with essentially no effect in other areas[12]. The global feedback systems are generally performed by the MInimisation des Carres des Distoution d'Orbite (MICADO) and the singular value decomposition (SVD) methods based on the response matrix or the harmonic correction method based on the harmonics. MICADO method is a modified least squares method based on the response matrix. The core idea is to minimize the number of correctors used while ensuring the calibration effect, which was first proposed by CERN in 1973[10]. The researchers at NSLS used to employ harmonic correction method for the closed orbit feedback in 1988[11], while the SVD method based on the response matrix is the most widely used. For instance, the researchers at Advanced Photon Source in Argonne used the SVD method of the response matrix for closed orbit correction[13]. The researchers at NSLS X-ray ring and the SPEAR also used a similar technique of SVD for global closed orbit correction in 1993[14]. The researchers at Taiwan Photon Source have been using SVD method to simulate closed orbit correction since the proposal stage in 2006[15]. The researchers at SSRF used the MML toolbox in Matlab to develop the Slow Orbit Feed Back (SOFB) system, based on SVD algorithm for the closed orbit feedback, in the beginning of accelerator commissioning in 2009[16]. To sum up, the commonly used closed orbit feedback methods are based on the MICADO and SVD algorithms, and the core of these two methods is the response matrix. The corresponding strength of correctors are calculated by the response matrix and the BPM data. These algorithms based on the response matrix are only linear mapping.

Nowadays, machine learning is more and more widely used in the accelerator field. For examples, researchers at SLAC replaced the traditional genetic and particle swarm algorithms with machine learning-based method to optimize the nonlinear problem of the storage ring and find multi-objective optimized Pareto front[17], and researchers at LBNL replaced the traditional feed forwards method with machine learning algorithms to counteract the insertion device gap or phase motion-induced perturbations on the ALS light source electron beam [18]. Furthermore, innovative polynomial neural networks for fitting beam dynamics parameters in accelerators are proposed by A. Ivanov and I. Agapov[19]. Their networks are ideally suitable for representing beam dynamics in accelerators. In addition, researchers from the Institute of High Energy Physics have proposed a new dynamic aperture (DA) prediction method based on machine learning which can reduce the computation cost of DA tracking by approximately one order of magnitude, while keeping sufficiently high evaluation accuracy[20].

In this paper, we explore the application of machine learning for a closed orbit feedback. We demonstrate how the application of machine learning allows for the closed orbit feedback relying only on the measured BPM data. This machine learning-based new method can provide precise results, while the residual values of the correctors' currents variations after correction can be almost ignored. This paper is organized with the following sections. Section 2 explains the theory of closed orbit feedback and machine learning; Section 3 presents the experimental research based on new method, including the experimental method, experimental results and analysis; Section 4 makes a brief summary of this paper.

## II. THEORY OF CLOSED ORBIT FEEDBACK AND MACHINE LEARNING

### A. Classical theory

There are 138 BPM in the storage ring of SSRF, and each BPM will measure a beam position. Therefore, we can define a vector $\Delta \boldsymbol{u}$ to represent the COD with the unit of meter and the dimension 138, $\Delta \boldsymbol{u} = (\Delta u_1, \Delta u_2, \cdots, \Delta u_m, \cdots, \Delta u_{138})^T$ where $u$ represents the orbit in the horizontal or vertical direction at the BPM. Then we define a vector $\boldsymbol{\theta}$ to represent the strength change of correctors with the unit of radian and the dimension 80, $\boldsymbol{\theta} = (\theta_1, \theta_2, \cdots, \theta_n, \cdots, \theta_{80})^T$ where $\theta$ represents the strength change of correctors in the horizontal or vertical direction. Each corrector causes a deflection angle depending on the strength of the corrector and the orbital variation is proportional to the deflection angle. With $i$ representing the number of the BPM and $j$ representing the number of the corrector, the scale factor of the COD at the $i$-th BPM caused by the strength of the $j$-th corrector can be written as[21]:

$$R_{ij} = \frac{\sqrt{\beta_i \beta_j}}{2 \sin(\pi \nu)} \cos(|\psi_i - \psi_j| - \pi \nu) + \frac{\eta_i \eta_j}{\alpha_c L_0} \tag{1}$$

where $\beta$ is the beta function, $\eta$ is the dispersion function, $\alpha_c$ is the momentum compact factor, $L_0$ is the circumference, $\psi$ is the phase and $\nu$ is the transverse working point. There are 138 BPM and 80 correctors so $R_{ij}$ can form a matrix $\boldsymbol{R}$ of 138 rows and 80 columns. This matrix $\boldsymbol{R}$ is also called response matrix. Relationship between COD and strength change of correctors is given by:

$$\Delta \boldsymbol{u} = \boldsymbol{R}\boldsymbol{\theta}. \tag{2}$$

Nevertheless, the response matrix is often not calculated by applying the formula in actual operation, but by applying the linear optics from closed orbit experimental measurements to obtain the machine response matrix. Based on the above analysis, the corrected closed orbit position can be obtained as:

$$\boldsymbol{U} = \Delta \boldsymbol{u} + \boldsymbol{R}\boldsymbol{\theta} \tag{3}$$

and the smaller the value, the better. Therefore, we can obtain the corresponding strength of correctors to correct the COD as:

$$\boldsymbol{\theta} = -\boldsymbol{R}^{-1} \Delta \boldsymbol{u}. \tag{4}$$

$\boldsymbol{\theta}$ needs to be solved by the above equations, but response matrix $\boldsymbol{R}$ is not a square matrix. It's the most common to use the SVD algorithm for approximate solution since we cannot solve inverse of a non-square matrix directly.

By the SVD algorithm we can solve the pseudo-inverse of response matrix $\boldsymbol{R}$. By using this method, response matrix $\boldsymbol{R}$ can be written by:

$$\boldsymbol{R} = \boldsymbol{u}\boldsymbol{s}\boldsymbol{v}^T \qquad (5)$$

and

$$\boldsymbol{R}^{-1} = \boldsymbol{v}\boldsymbol{s}^{-1}\boldsymbol{u}^T \qquad (6)$$

where $\boldsymbol{u}$ is a 138×138 unitary matrix, $\boldsymbol{s}$ is a 138×80 diagonal matrix and $\boldsymbol{v}$ is an 80×80 unitary matrix. The elements on the main diagonal of the matrix $\boldsymbol{s}$ are also called singular values. The number of singular values $N$ is manually selected. Finally, the strength of correctors can be written as:

$$\boldsymbol{\Theta} = -\boldsymbol{v}\boldsymbol{s}^{-1}\boldsymbol{u}^T \Delta \boldsymbol{u}. \qquad (7)$$

The above is the classic orbit feedback principle.

However, the classic orbit feedback method also has its shortcomings. For example, the measurement of the response matrix will consume valuable machine study time and could not be updated in the real time. The drift of various parameters is generated with the passage of time during the daily operation. The change of insertion devices will also cause the change of storage ring status. The response characteristics of the closed orbit are changing all the time. Therefore, the response matrix must be measured and updated periodically.

For the closed orbit feedback method, commonly used in worldwide light sources, the SVD algorithm based on MATLAB platform can quickly calculate the pseudo-inverse matrix. However, the BPM error, BPM measurement accuracy and the selection of the number of singular values will affect the results of closed orbit feedback. The selection of the number of singular values generally has a great impact on the results of the correction. We usually choose to discard relatively small singular values, as they are more susceptible to noise. If the quantity of singular values in use is not enough, the result may not be unique, which is the case of under-constraint. After long-term use of SVD algorithm for closed orbit feedback, the strengths of some correctors will drift in one direction or even saturate due to under-constraint, which may finally lead to the crash of the system. What's more, the traditional closed orbit feedback methods are based on a given response matrix for calculation and correction, which are only linear mapping and does not fully and accurately reflect the machine characteristics.

### B. Machine learning theory

The new idea about closed orbit feedback is the application of machine learning, which allows for the closed orbit feedback relying only on the measured BPM data. In recent years, with the development of machine learning and neural networks, Data-Driven method has been used in many different fields[22]. It does not require any assumptions or prior knowledge. Through the neural networks model, the method provides a completely data-based information mining tool.

Machine learning uses neural networks to fit data composed of many nodes, which can solve complex problems. These nodes, also known as neurons, are the most basic components of neural networks, proposed by McCulloch and Pitts in 1943[23]. It can accept inputs from other nodes or external data and then calculate an output. The output data is obtained by weighting the input data and adding the bias finally calculating by the activation function. Fig. 1 shows the schematic illustration of a node. The output data of a node can be written as:

$$Y = f(\sum_n w_i X_i + b) \qquad (8)$$

where $X$ is the input data, $w$ is the weight, $b$ is the bias, $f$ is the activation function and $Y$ is the output data. For dense neural networks, each layer is composed of several nodes or a single node, and all nodes in the previous layer connect to those in the next layer. The layers of a neural network can be classified into the input layer, hidden layers and output layer. The input data enter the input layer, and then enter the hidden layers for calculation. The calculation results of the hidden layers are transmitted to the output layer for output. This is the forward propagation of the signals from input to output. In the neural network, the labels represent the true conclusions of the predicted things and can be understood as the correct answers. The features are the inherent traits of things, which can be understood as the basis for obtaining true conclusions and they are usually used as input variables for the neural network model. The output data in a good neural network model should be as close as possible to the labels corresponding to the input features. For different types of data, people usually set different loss functions to measure the errors of output signals and labels. The loss function also represents the performance of machine learning.

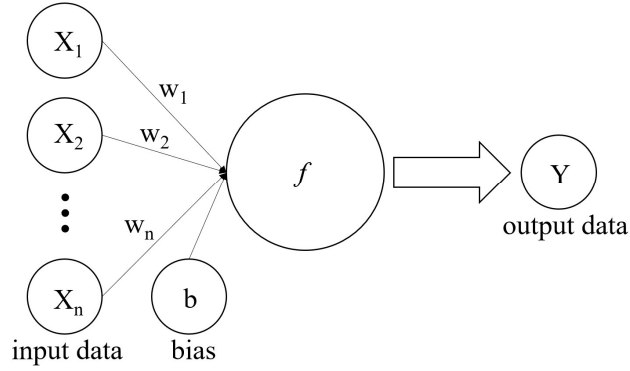

FIG. 1. Schematic illustration of a node.

For dense neural networks, the input data of the nth layer can be written as:
$$I^{(n)} = w^{(n)} \cdot o^{(n-1)} + b^{(n)} \tag{9}$$

where $o^{(n-1)}$ is the output data of the (n-1)-th layer, $w^{(n)}$ and $b^{(n)}$ are the weights and biases of the nth layer respectively. Equation (9) can be further written as:

$$I^{(n)} = \begin{bmatrix} I_1^{(n)} \\ \vdots \\ I_i^{(n)} \end{bmatrix} = \begin{bmatrix} w_{1,1}^{(n)} & \cdots & w_{1,j}^{(n)} \\ \vdots & \ddots & \vdots \\ w_{i,1}^{(n)} & \cdots & w_{i,j}^{(n)} \end{bmatrix} \cdot \begin{bmatrix} o_1^{(n-1)} \\ \vdots \\ o_j^{(n-1)} \end{bmatrix} + \begin{bmatrix} b_1^{(n)} \\ \vdots \\ b_i^{(n)} \end{bmatrix} \tag{10}$$

where $i$ is the number of nodes in the nth layer, $j$ is the number of nodes in (n-1) th layer. After calculating by the activation function, the output data can be obtained:

$$o^{(n)} = f(I^{(n)}) = \begin{bmatrix} f(I_1^{(n)}) \\ \vdots \\ f(I_i^{(n)}) \end{bmatrix} = \begin{bmatrix} o_1^{(n)} \\ \vdots \\ o_i^{(n)} \end{bmatrix} \tag{11}$$

where $f$ is the activation function. If the nth layer is output layer, the $o^{(n)}$ is the output data of the neural network. Then the loss functions can be used to measure the errors between output data and labels. The model is optimized by reducing the value of this loss function. The optimization process is then to find a set of $w$ and $b$ to minimize the loss function.

The most commonly used optimization method is called error back propagation method[24], in which errors start from the output layer and move backward until they reach the right next hidden layer to the input layer. In this process, the data still flows through the connecting lines and the weights are multiplied. The gradient of each node in the network is calculated using the chain rule. This process is called gradient descent optimization[25]. The formula of the gradient descent optimization can be written by:

$$w^{(n)} = w^{(n)} - \alpha \frac{\partial L}{\partial w^{(n)}} \tag{12}$$

and

$$b^{(n)} = b^{(n)} - \alpha \frac{\partial L}{\partial b^{(n)}} \tag{13}$$

where $L$ is the loss function, $\alpha$ is the learning rate. Learning rate is used to control the range of weights adjustment. The principle of this optimization method is that the direction of the gradient is the fastest rising direction of the objective function in the given point. Therefore, iterating along the reverse direction of the gradient, the decrease of the loss function can be achieved. After the training is completed, the neural networks can predict the corresponding result data when the data similar to feature is inputted. Fig. 2 shows the schematic illustration of dense neural networks.

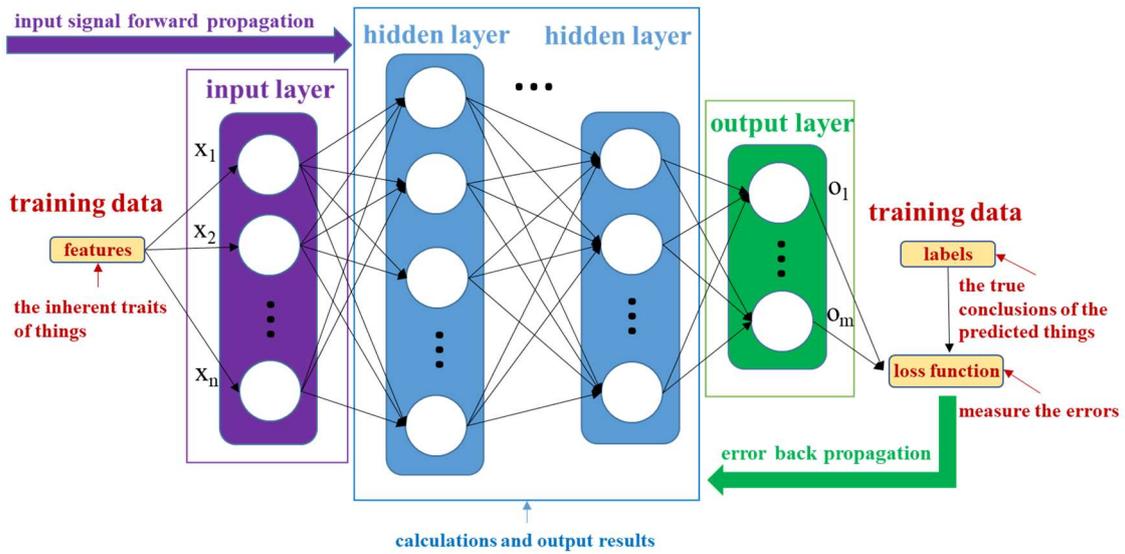

Fig. 2. Schematic illustration of a dense neural network. (x represents the input signal, o represents the output signal).

We can then use the trained model to reach our goals in terms of physics. Our network is used to obtain the corresponding strength of correctors and RF frequency after inputting BPM data, so we use BPM data as the input data and use correctors strength and RF frequency as the output data. Fig. 3 show the schematic illustration of the machine learning for a closed orbit feedback. The machine learning-based method actually establishes the nonlinear mapping relationship between the COD and the correctors' currents. It's more like a "black box" instead of relying on the specific response matrix. The strength of correctors can be directly predicted from the measured BPM data by the machine learning-based method. This process bypasses specific physical model and can also save the measurement time of the response matrix. Moreover, the neural network is a nonlinear mapping, which can model the machine more accurately than the linear mapping such as traditional SOFB.

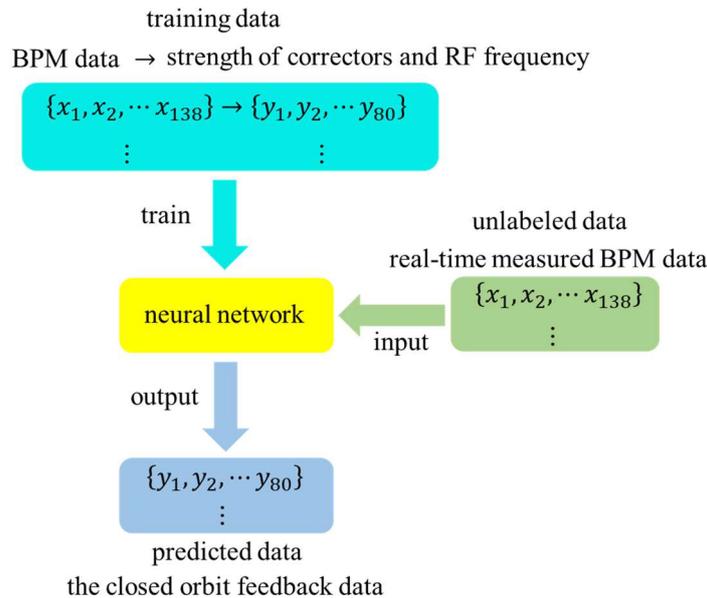

FIG. 3. Schematic illustration of the machine learning for a closed orbit feedback.

## III. EXPERIMENTAL RESEARCH ON MACHINE LEARNING

### A. Methods

#### 1. Building the neural networks model

Selecting machine learning module is the first step. Our machine learning algorithm is programmed by using Python. Keras is used as a deep learning module and it is a high-level API based on TensorFlow[26]. TensorFlow, based on dataflow programming, is widely used in the programming of various machine learning algorithms.

Then we start building the neural networks model, using root mean squared error (RMSE) as the loss function and mean absolute error (MAE) as the metrics function. The model is trained using the error back propagation method and employing the Stochastic Gradient Descent(sgd) optimizer[27]. Learning rate is the step size of weights adjustment. Too high learning rate leads to failure of converging and the result will hover around the optimal value for the network. However, if the learning rate is too low, the network will converge very slowly. In our neural networks the learning rate is set at 0.01, and the performance of different learning rates will be discussed.

The topological structure of the neural networks is particularly important. The deeper neural networks with more hidden layer and the wider neural networks with more nodes in each layer the better it can fit the training data. However, a large model is prone to over-fitting and requires more computational resources with longer time in the training process. Based on our data type, a fully connected neural network with three hidden layers was finally selected and each hidden layer has 100 nodes. We also tried deeper networks and more nodes but there was no significant improvement of performance. The current choice can compromise between appropriate computing resources and good performance. We use rectified linear unit (ReLU) function as activation function in nodes to simulate the process of neurons respondence according to signal strength[28]. The definition of rectified linear unit function is given by $f(x) = max\,(0,x)$. Rectified linear unit function has the characteristics of unilateral suppression (part of nodes are set to 0) and relatively broad excitation boundary. Rectifying neurons can create sparse representations with true zeros and have yielded equal or better performance than hyperbolic tangent networks[29]. In our model its learning speed is faster than other activation functions and the accuracy is also higher. A small $L_2$ regularization parameter with $\lambda = 10^{-4}$ was also used to reduce overfitting. $L_2$ regularization, also known as weight decay, is widely used in various models[30]. This method decay weights by adding the sum of squared weights multiplied with the regularization parameter to the original loss function, which is helpful to improve the generalizability of the model. The next step is to initialize the weights of the neural network model. Weights initialization refers to the process of initializing the weights and biases assigned to each node before the neural network model is trained. The purpose of weights initialization is to prevent gradient explosion or gradient disappearance in the forward propagation process of deep neural networks, which has an impact on the convergence speed and performance of the model[31]. Gradient explosion or gradient disappearance will lead to a very large or very small derivative of the neural networks in the back propagation training, which makes the weights too heavy or unable to be updated. We use the normal distribution to initialize the weights, which can improve the convergence speed and performance of our model.

#### 2. Training data acquisition and preprocessed

Training data is crucial to train the neural network and the quality of data determines the performance of the model. In SSRF, the Experimental Physics and Industrial Control System (EPICS) is used to collect data. We need the BPM data and the corresponding strength of the correctors. We collected 40 minutes of BPM data and the corresponding strength of the correctors with a sampling interval of three seconds. These data cannot be used directly. They need to be preprocessed. For the collected data grouped by BPM serial number and corrector serial number, each BPM and each corrector data is a separate group. For each group of data, we require the variations by $dx_i = x_i - x_0$ where $x_0$ represents the data at the initial moment. For different BPM and correctors, the magnitude of the variations may vary greatly because of their different characteristics and states. The largest BPM in our data exceeds 4 mm and the smallest BPM is only about 10 μm. When these two data are directly input into the neural network model for training without preprocessing, the larger value will occupy a heavier weight, while the smaller value will be almost ignored during the training process. This approach will seriously affect the training effect of the neural network model. Hence, the training data of neural networks usually need to be standard scaler. The formula for standard scaler is as follows: $x' = \frac{x-\mu}{\sigma}$ where $\mu$ is the mean of the samples and $\sigma$ is the standard deviation of the samples. Although normalization is also possible, normalization is the mapping of data to the interval [0,1] or [-1,1] based on the extreme values. Unlike normalization, standard scaler is related to the overall sample distribution and each sample can have an impact on standard scaler so the sample spacing can be better maintained. Through standard scaler, each sample can be scaled to the state of mean 0 and variance 1 without changing the distribution of the original data to make sure the samples among different dimensions are numerically comparable. In this way, all training data can be reduced to the same range to train the model, which will minimize the effect of large variance data, improve the generalization ability of our model and optimize the training results.

### B. Results and analysis

The training takes less than two minutes on a single laptop-class CPU R9-5900HX (core frequency 3.3 GHz). The 800 sets of data were divided into 700 sets of training data and 100 sets of test data. The test data was not involved in the training but only as validation. Fig. 4 shows the relationship between the MAE of training data and test data and the number of epochs and the effects of different learning rates are compared. It can be seen from the figure that the MAE of the training data can converge and is basically stable after 100 epochs of the training. When the learning rate is reduced to 0.005, the MAE converge more slowly. When the learning rate is increased to 0.15, the convergence speed does not improve significantly but the fluctuation of MAE becomes larger. Therefore, in our neural networks the learning rate is finally set at 0.01. In addition, no obvious overfitting phenomenon is observed.

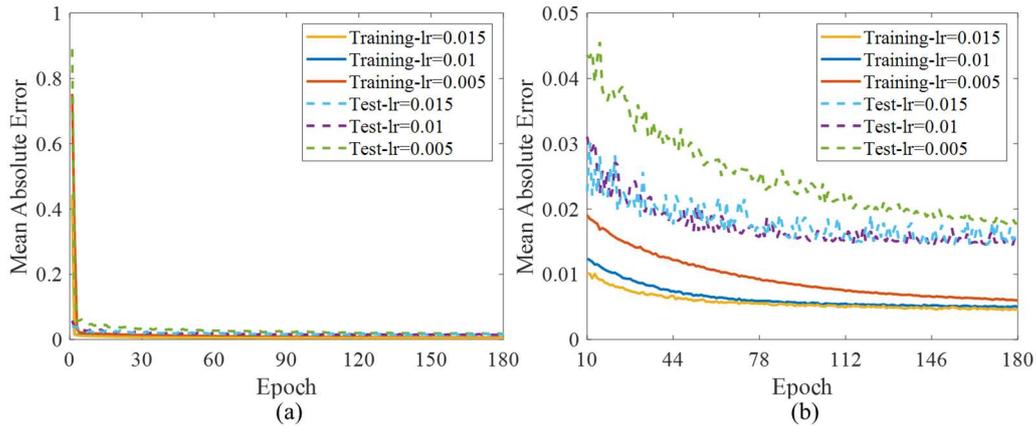

FIG. 4. Mean absolute errors of training data and test data. (b) is the blowup of (a).

After that, we carried out an experiment of the machine learning-based method at SSRF. The experiment is divided into four parts, namely, (a) the RF frequency feedback experiment, (b) the joint feedback experiment including RF frequency and correctors, (c) the feedback experiment of SOFB system and finally (d) the long-term operation experiment of the machine learning-based closed orbit feedback. The reason for this arrangement is that the RF frequency feedback is relatively simple so it is performed before the joint feedback experiment. The experiment for the SOFB system is for comparison.

*1. RF frequency feedback*

The RF frequency feedback experiment was carried out under top-up operation (201~202mA). We turned on the RF frequency feedback program and set five-second feedback at a time. Then we manually reduced 30 Hz from the original RF frequency. At this moment, the orbit changed and we observed that the RF frequency feedback program could correct the RF frequency back according to the variations of orbit. Fig. 5 shows the orbit and RF frequency variations. The RF frequency was changed a bit at a time, because the RF frequency of SSRF storing ring can only it is limited to be changed by a maximum of 10 Hz at a time. As can be seen from figure 3(b), the machine learning program can correct the RF frequency back to the original value within 10 seconds or so.

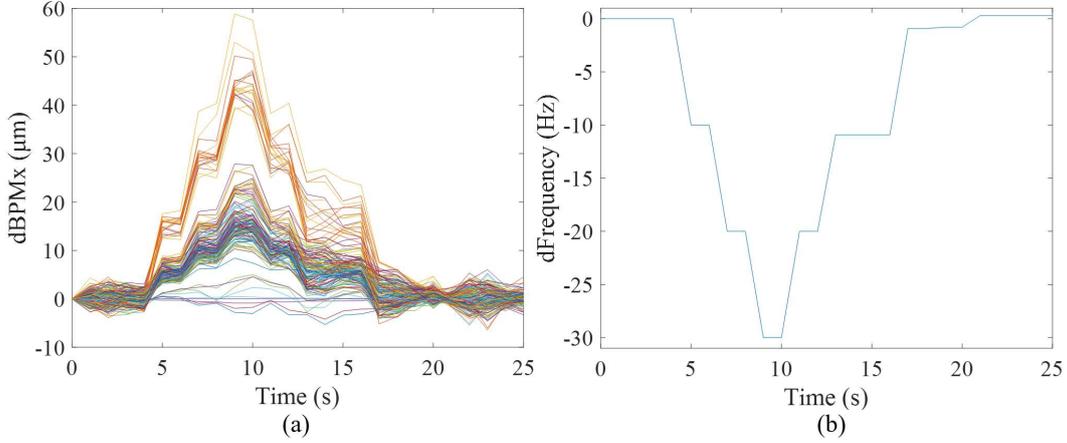

FIG. 5. The variations of horizontal orbit and RF frequency. (a) the variations of horizontal orbit and (b) the variations of RF frequency.

### *2. Joint feedback of RF frequency and correctors*

The experiment was also carried out under top-up operation (201~202mA). We turned on the machine learning-based feedback program and set five-second feedback at a time. We gave a random disturbance to the horizontal plane and the vertical plane correctors (HCM and VCM) respectively and meanwhile the RF frequency was added by 15 Hz. At this moment, the orbit changed and we observed whether the machine learning-based feedback program could correct the HCM, VCM and RF frequency back according to the variations of orbit. Fig. 6 and Fig. 7 show the orbit, correctors and RF frequency variations. As can be seen from the figures, after about three feedback loops, the machine learning-based method can correct the orbit back to be near its previous level, and the currents of HCM and VCM and the RF frequency are also corrected back to around the initial values. We use the horizontal and vertical orbit RMS and peak-to-peak value to measure the experimental results, as listed in Table 1. After correction these parameters are closed to the initial value.

TABLE I. The RMS and peak-to-peak value of all BPM in the horizontal plane and vertical plane with the machine learning-based feedback program turned on.

| The experimental result | RMS(H) (μm) | RMS(V) (μm) | Peak-to-peak value (H) (μm) | Peak-to-peak value(V) (μm) |
|---|---|---|---|---|
| Before correction | 1.648 | 0.856 | 7.698 | 4.253 |
| Random disturbance | 384.4 | 192.3 | 1821 | 918.5 |
| After correction | 1.907 | 1.162 | 9.658 | 5.65 |
| Deviation (after-before) | 0.259 | 0.306 | 1.96 | 1.397 |

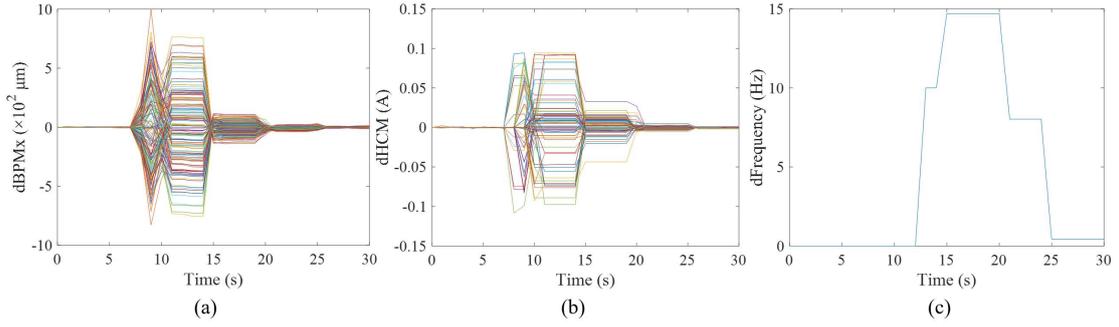

FIG. 6. The variations of horizontal orbit, HCM strength and RF frequency with the machine learning-based feedback program turned on. (a) the variations of horizontal orbit, (b) the variations of HCM strength and (c) the variations of RF frequency.

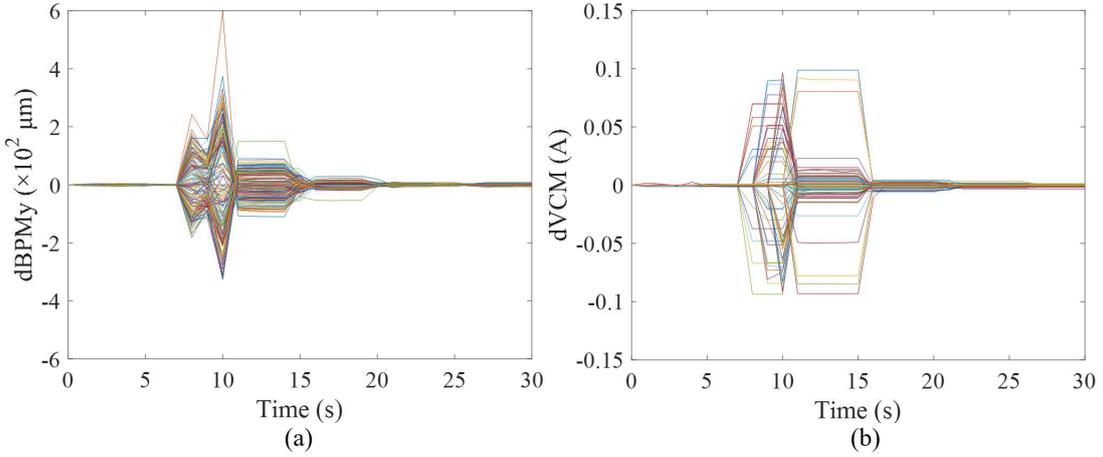

FIG. 7. The variations of vertical orbit and VCM strength with the machine learning-based feedback program turned on. (a) the variations of vertical orbit and (b) the variations of VCM strength.

*3. SOFB system feedback*

The traditional SVD-based SOFB experiment is also carried out under similar conditions with three-second feedback at a time. Fig. 8 and Fig. 9 show the orbit, correctors and RF frequency variations. As can be seen from the figures, after about ten feedback loops, although the SOFB can correct the orbit back to be near its previous level, the currents of HCM and VCM have significant residual values and the correction effect of RF frequency is not meeting our expectation. The issue about residual values will be discussed. We also use the horizontal and vertical orbit RMS and peak-to-peak value to measure the experimental results, as listed in Table 2.

TABLE II. The RMS and peak-to-peak value of all BPM in the horizontal plane and vertical plane with the SOFB system turned on.

| The experimental result | RMS(H) (μm) | RMS(V) (μm) | Peak-to-peak value (H) (μm) | Peak-to-peak value(V) (μm) |
|---|---|---|---|---|
| Before correction | 2.384 | 1.137 | 13.59 | 4.724 |
| Random disturbance | 188.4 | 263.6 | 914.4 | 1004 |
| After correction | 5.487 | 2.448 | 32.29 | 17.41 |
| Deviation (after-before) | 3.103 | 1.311 | 18.70 | 12.69 |

According to the above data, it can be seen that although SOFB can also correct the orbit, the SVD-based SOFB is not as effective as the machine learning-based method. The machine learning-based method is better than the SOFB based on the response matrix in terms of convergence and convergence speed. What's more, the significant residual values in the SVD-based SOFB may cause the correctors' currents to saturate during long-term operation, resulting in system collapse. The reason for the residual values is that the number of singular values used in the SVD method is not very appropriate when using the SOFB for orbit feedback, resulting in an under-constrained solution that is not unique. We can discuss the following situation. When there is an orbital error that needs to be corrected with the corrector 'A', the SOFB uses many correctors together to correct the error instead of just using the corrector 'A'. In the same case, when using the machine learning-based method for correction, it can accurately locate the corrector 'A' and correct the orbit, which is an obvious advantage. Since the SOFB leaves the residual values of the correctors' currents variations in the feedback process, it is necessary to manually adjust the correctors' currents at intervals to prevent saturation of the correctors' currents during daily operation. This is not in need for the new machine learning-based method, which is a significant advantage.

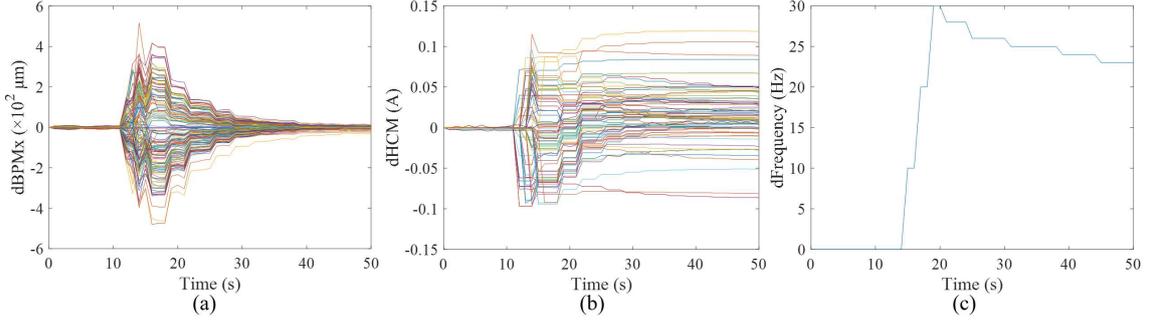

FIG. 8.  The variations of horizontal orbit, HCM strength and RF frequency with the SOFB system turned on. (a) the variations of horizontal orbit, (b) the variations of HCM strength and (c) the variations of RF frequency.

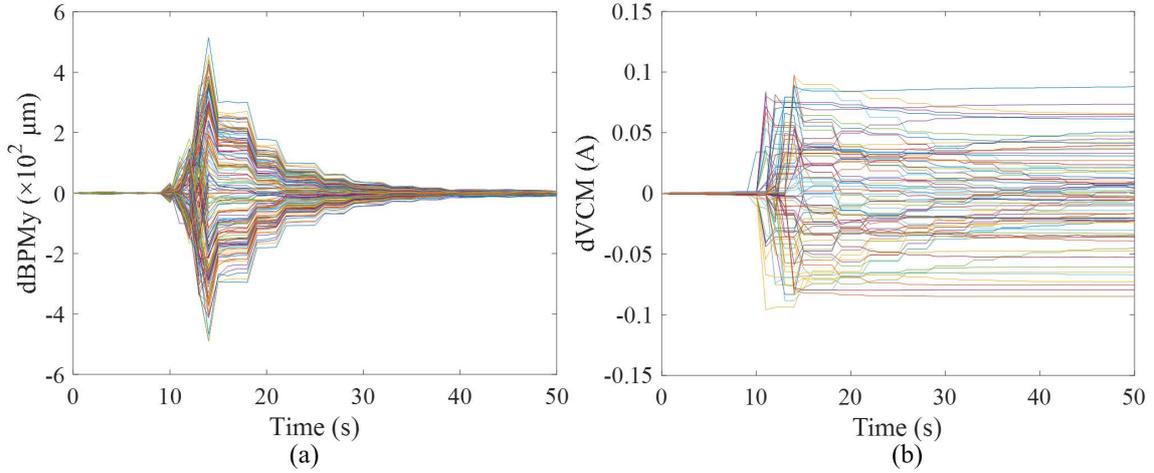

FIG. 9.  The variations of vertical orbit and VCM strength with the SOFB system turned on. (a) the variations of vertical orbit and (b) the variations of VCM strength.

*4. Long-term operation*

We did a long-term operation experiment under top-up operation (201~202 mA), and it lasted 16 hours from 3:00 pm to 7:00 am in the next morning. Fig. 10 and Fig. 11 show the orbit, RMS orbit, correctors and RF frequency variations. During this process the RMS of orbit variations at all BPM are basically below 15 μm and 8 μm in the horizontal and vertical plane respectively. In addition, with the machine learning-based feedback program turned on, the maximum orbit drift is about 40 μm and 20 μm in the horizontal and vertical plane respectively and without the orbit feedback, the daily orbit drift can reach a few hundred microns. As can be seen, the orbit feedback can reduce the orbit drift by an order of magnitude. What's more, two clear inflection points can be seen on the graph of the RF frequency variations, which correspond to two inflection points of the environment temperature in a day. A change of the circumference due to thermal expansion will affect a change of the orbit. In general, the RF frequency should be changed in order to keep the orbit stable. Therefore, the RF frequency should be reduced when the temperature rises and vice versa, which is exactly the same as the result given by our machine learning-base closed orbit feedback. In addition, we found a fluctuation in the vertical orbit at the third hour with a maximum of about 20 μm, which is supposed to be related to the unstable state of the machine. We believe that it was not an error of our method that caused this situation, since similar fluctuations were not observed in the horizontal plane and did not reappear in subsequent long-term tracking.

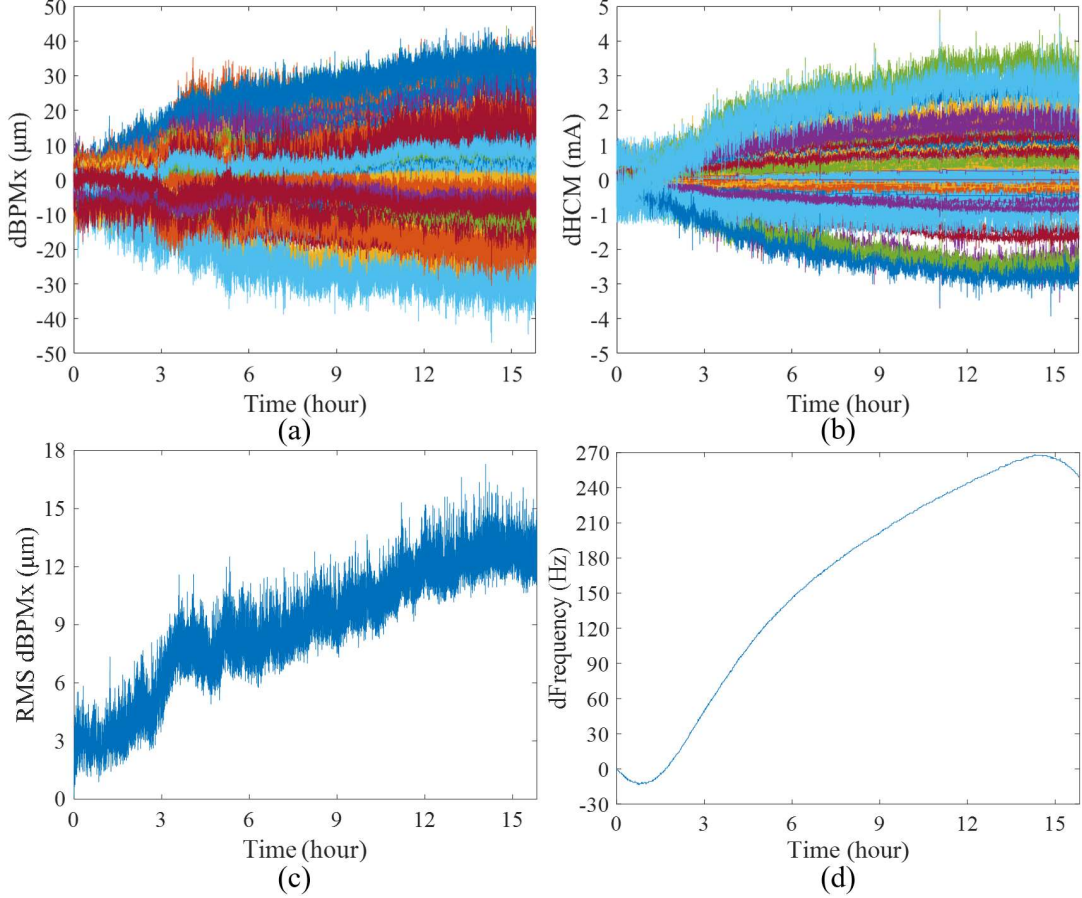

FIG. 10. The variations of horizontal orbit, HCM strength, horizontal orbit RMS and RF frequency with the machine learning-based feedback program turned on. (a) the variations of horizontal orbit, (b) the variations of HCM strength, (c) the variations of horizontal orbit RMS and (d) the variations of RF frequency.

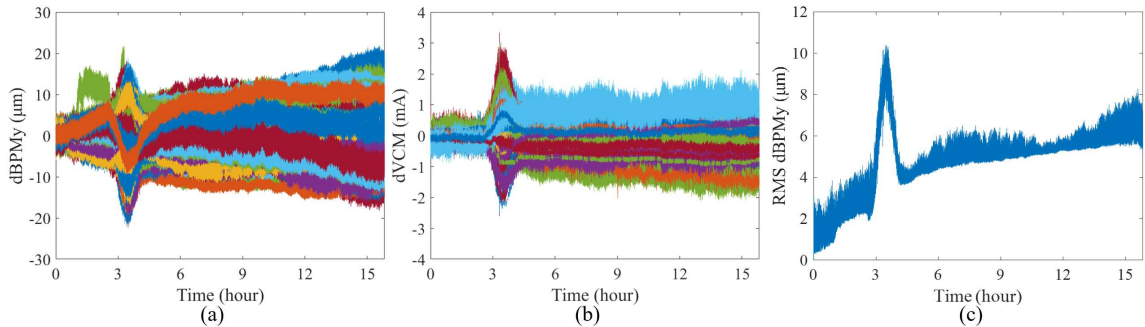

FIG. 11. The variations of vertical orbit, VCM strength, vertical orbit RMS and RF frequency with the machine learning-based feedback program turned on. (a) the variations of vertical orbit, (b) the variations of VCM strength and (c) the variations of vertical orbit RMS.

## IV. Conclusion

This paper investigates the machine learning-based new method and the experiments are done at the SSRF storage ring. Our study demonstrates that this new machine learning-based method can be used for a closed orbit feedback. We compare the effects of the new method and the traditional SOFB orbit feedback through experiments. The experimental results show that the closed orbit feedback based on machine learning is better than the SOFB based on the response matrix in terms of convergence and convergence speed. One important difference is that the traditional SOFB system leaves the residual values of the correctors' currents variations during the feedback process, while this situation does not happen with the machine learning-based method. This new method can accurately pinpoint and correct the correctors. We also did experiments to demonstrate that this machine learning-based feedback program remains effective over a long period of time and the achieved orbit stability meets our expectations. In addition, valuable machine study time for measuring the response matrix can be saved if the new method is adopted. This new method provides a new way for orbit feedback of light sources and its feasibility is demonstrated, which is of great significance for improving the orbit stability of the storage ring in daily operation.

## Acknowledgments

This work is supported by National Natural Science Foundation of China (No. 11975298) and the Youth Innovation Promotion Association of CAS (No.2020287).